\documentclass[twocolumn,showpacs,preprintnumbers,amsmath,amssymb]{revtex4}

\usepackage{graphicx}
\usepackage{dcolumn}
\usepackage{bm}

\begin{document}

\preprint{APS/123-QED}

\title{Magnetic susceptibilities in a family of ${\bm S}$\,{=}\,1/2 Kagom\'{e} antiferromagnet}

\author{T. Ono$^1$}
\email{o-toshio@lee.phys.titech.ac.jp.}
\author{K. Morita$^1$}
\author{M. Yano$^1$}
\author{H. Tanaka$^1$} 
\author{K. Fujii$^2$}
\author{H. Uekusa$^2$}
\author{Y. Narumi$^3$}
\author{K. Kindo$^4$}
\affiliation{
$^1$Department of Physics, Tokyo Institute of Technology, Meguro-ku, Tokyo 152-8551, Japan\\
$^2$Department of Chemistry and Materials Science, Tokyo Institute of Technology, Meguro-ku, Tokyo 152-8551, Japan\\
$^3$Institute for Material Research, Tohoku University, Aoba-ku, Sendai 980-8577, Japan\\
$^4$Institute for Solid State Physics, The University of Tokyo, Kashiwa, Chiba 277-8581, Japan
}

\date{\today}

\begin{abstract}
Hexagonal antiferromagnets Cs$_2$Cu$_3$MF$_{12}$ (M\,=\,Zr, Hf and Sn) have uniform Kagom\'{e} lattices of Cu$^{2+}$ with $S\,{=}\,1/2$, whereas Rb$_2$Cu$_3$SnF$_{12}$ has a $2a\,{\times}\,2a$ enlarged cell as compared with the uniform Kagom\'{e} lattice. The crystal data of Cs$_2$Cu$_3$SnF$_{12}$ synthesized first in the present work are reported. We performed magnetic susceptibility measurements on this family of Kagom\'{e} antiferromagnet using single crystals. In the Cs$_2$Cu$_3$MF$_{12}$ systems, structural phase transitions were observed at $T_{\rm t}\,{=}\,225$ K, 172 K  and 185 K for M\,=\,Zr, Hf and Sn, respectively. The magnetic susceptibilities observed for $T\,{>}\,T_{\rm t}$ are almost perfectly described using theoretical results obtained by exact diagonalization for the 24-site Kagom\'{e} cluster with $J/k_{\rm B}\,{=}\,244$ K, 266 K and 240 K, respectively. Magnetic ordering accompanied by the weak ferromagnetic moment occurs at $T_{\rm N}\,{=}\,23.5$ K, 24.5 K and 20.0 K, respectively. The origins of the weak ferromagnetic moment should be ascribed to the lattice distortion that breaks the hexagonal symmetry of the exchange network for $T\,{<}\,T_{\rm t}$ and the Dzyaloshinsky-Moriya interaction. 
Rb$_2$Cu$_3$SnF$_{12}$ is magnetically described as a modified Kagom\'{e} antiferromagnet with four types of neighboring exchange interaction. Neither structural nor magnetic phase transition was observed in Rb$_2$Cu$_3$SnF$_{12}$. Its magnetic ground state was found to be a spin singlet with a triplet gap. Using exact diagonalization for a 12-site Kagom\'{e} cluster, we analyzed the magnetic susceptibility and evaluated individual exchange interactions. The causes leading to the different ground states in Cs$_2$Cu$_3$SnF$_{12}$ and Rb$_2$Cu$_3$SnF$_{12}$ are discussed.
\end{abstract}

\maketitle


\section{Introduction}
The Heisenberg Kagom\'{e} antiferromagnet (HKAF) has been attracting considerable attention from the viewpoints of frustration and quantum effects \cite{Harrison,ML}. In the classical spin model, three spins on a local triangle form a $120^{\circ}$ structure as in the case of the Heisenberg triangular antiferromagnet (HTAF). However, different from the HTAF, the classical HKAF has infinite ground states in the case of the nearest neighbor exchange interaction because of the flexibility of configuration of neighboring two $120^{\circ}$ structures. When the next-nearest neighbor (NNN) interaction is switched on, the so-called $q\,{=}\,0$ or $\sqrt{3}\,{\times}\,\sqrt{3}$ structure is stabilized in accordance with the sign of the NNN interaction \cite{Harris}. The HKAF with the nearest neighbor interaction has been actively studied theoretically. It was predicted that for classical and large spins, thermal and quantum fluctuations stabilize the $\sqrt{3}\,{\times}\,\sqrt{3}$ structure without the help of the NNN interaction \cite{Chubukov,Reimers}, whereas for small spin values as $S\,{=}\,1/2$ or 1, strong quantum fluctuation leads to the disordered ground state \cite{Zeng1,Sachdev,Chalker,Elstner,Nakamura,Hida1,Hida2}. Careful analysis and numerical calculation for the $S\,{=}\,1/2$ case revealed that triplet excitations are gapped, but there exists the continuum of low-lying singlet states below the triplet gap \cite{Lecheminant,Waldtmann,Mila,Syromyatnikov}. The magnitude of the triplet gap was estimated to be of the order of one-tenth of $J$, where the exchange constant $J$ is defined as ${\cal H}_{\rm ex}\,{=}\,\sum_{\langle i,j\rangle} J_{ij}\,{\bm S}_i{\cdot}{\bm S}_j$. Valence-bond crystal represented by a periodic arrangement of singlet dimers has been proposed as the ground state of $S\,{=}\,1/2$ HKAF \cite{Nikolic,Budnik,Singh1,Singh2}. 
An other theory based on the resonating valence bond or the gapless critical spin liquid has also been proposed \cite{Mambrini,Hastings,Ryu,Ran,Hermele}.

For $S\,{=}\,1/2$ HKAF, numerical calculation demonstrated that the magnetic susceptibility has a rounded maximum at $T_{\rm max}\,{\simeq}\,(1/6)J/k_{\rm B}$ and decreases toward zero with decreasing temperature \cite{Elstner,Misguich2}. The magnetic susceptibility obeys roughly the Curie-Weiss law in a wide temperature range for $T\,{>}\,(1/2)J/k_{\rm B}$. However, for $T\,{<}\,2J$, the absolute value of the effective Weiss constant ${\Theta}_{\rm eff}$ is given as $|{\Theta}_{\rm eff}|\,{\simeq}\,2J$, which is twice as large as $|{\Theta}|\,{=}\,J$ for molecular field approximation \cite{Rigol1,Rigol2}. These unusual features of the magnetic susceptibility have not been sufficiently verified.

The experimental studies on the HKAF have first been concentrated on jarosites with large spin values \cite{Takano1,Takano2,Townsend,Maegawa,Wills1,Inami1,Wills2,Inami2,Wills3,Morimoto,Grohol1,Nishiyama,Grohol3}. The chemical formula of the jarosites is expressed as AM$_3$(OH)$_6$(SO$_4$)$_2$, where A is a monovalent ion such as K$^+$ and M\,=\,Fe$^{3+}$\,($S\,{=}\,5/2$) or Cr$^{3+}$\,($S\,{=}\,3/2$). In the many jarosites, the ordered state with the $q\,{=}\,0$ structure has been observed contrary to the theoretical prediction \cite{Townsend,Inami1,Inami2,Grohol3}. The $q\,{=}\,0$ structure was also observed in hexagonal tungsten bronze (HTB)-type FeF$_3$ with the Kagom\'{e} layer \cite{Leblanc}. The $q\,{=}\,0$ structure observed in the jarosites and HTB-FeF$_3$ can be ascribed to the antisymmetric interaction of the Dzyaloshinsky-Moriya (DM) type with alternating ${\bm D}$ vectors \cite{Grohol3,Elhajal}. In iron jarosite with A\,=\,D$_3$O$^+$, spin freezing behavior was observed \cite{Wills4,Fak}. ($m$-MPYNN)$\cdot$BF$_4$$\cdot$$\frac{1}{3}$(acetone) is known as the composite $S\,{=}\,1$ HKAF with the gapped singlet ground state \cite{Wada}. In this organic system, two radical spins coupled through strong ferromagnetic interaction form the $S\,{=}\,1$ state, and the composite spins are placed on a modified Kagom\'{e} lattice \cite{Kambe}.

In contrast to theoretical studies, the experimental studies on the $S\,{=}\,1/2$ HKAF are limited. Experiments were performed on Cu$_3$V$_2$O$_7$(OH)$_2$\,$\cdot$\,2H$_2$O \cite{Hiroi}, [Cu$_3$(titmb)$_2$(CH$_3$CO$_2$)$_6$]\,$\cdot$\,H$_2$O\,\cite{Honda} and $\beta$-Cu$_3$V$_2$O$_8$ \cite{Rogado} that have a Kagom\'{e} or related lattice. Because the Kagom\'{e} net is distorted into an orthorhombic form in Cu$_3$V$_2$O$_7$(OH)$_2$\,$\cdot$\,2H$_2$O and buckled like a staircase in $\beta$-Cu$_3$V$_2$O$_8$, the exchange network is fairly anisotropic. For [Cu$_3$(titmb)$_2$(CH$_3$CO$_2$)$_6$]\,$\cdot$\,H$_2$O, the nearest neighbor exchange interaction is ferromagnetic \cite{Narumi}. The above-mentioned exotic ground state has not been observed in these systems.

Shores {\it et al.} \cite{Shores} reported the synthesis and magnetic properties of herbertsmithite ZnCu$_3$(OH)$_6$Cl$_2$ that has the uniform Kagom\'{e} lattice. Since then, herbertsmithite has been attracting considerable attention \cite{Mendels,Helton,Bert,Rigol1,Rigol2,Misguich2,Lee,Imai,Olariu,Vries,Zorko}. No magnetic ordering occurs down to 50 mK because of strong spin frustration \cite{Mendels}. However, the magnetic susceptibility exhibits a rapid increase at low temperatures, which was ascribed to idle spins produced by the intersite mixing of $5\,{\sim}\,10$ \% between Cu$^{2+}$ and Zn$^{2+}$ \cite{Helton,Bert,Rigol2,Misguich2}. NMR experiments on uniaxially aligned and $^{17}$O-enriched powder samples revealed that the Knight shift proportional to the intrinsic magnetic susceptibility of the host lattice exhibits a broad maximum at $50\,{-}\,100$ K, and decreases with decreasing temperature \cite{Imai,Olariu}. However, owing to the lattice disorder, the intrinsic magnetic susceptibility of herbertsmithite is still unclear.

\begin{figure}[t]
	\includegraphics[scale =0.8]{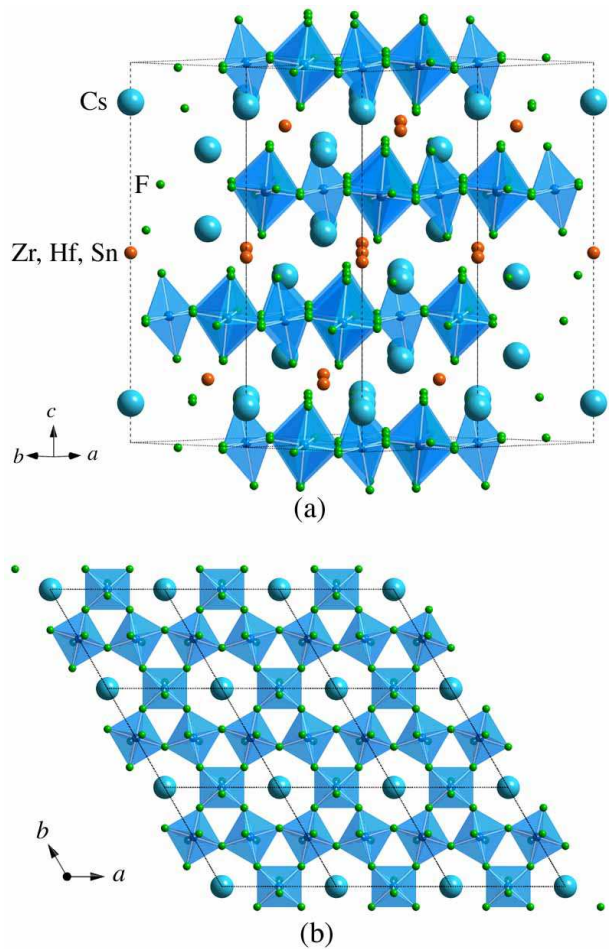}
\caption{(a) Crystal structure of Cs$_2$Cu$_{3}$MF$_{12}$ (M=Zr, Hf and Sn) viewed along the $[1, 1, 0]$ direction and (b) its projection onto the $c$ plane. Shaded octahedra represent CuF$_6$ octahedra. Dotted lines denote the unit cell.}
 \label{fig:structure1}
 \end{figure} 
 
Cupric compound Cs$_2$Cu$_3$MF$_{12}$ (M\,=\,Zr and Hf) is a new family of $S\,{=}\,1/2$ HKAF, which has a uniform Kagom\'{e} lattice at room temperature \cite{Mueller,Yamabe}. M\"{u}ller and M\"{u}ller \cite{Mueller} synthesized Cs$_2$Cu$_3$ZrF$_{12}$ and Cs$_2$Cu$_3$HfF$_{12}$ and determined the crystal structures at room temperature. These compounds are crystallized in hexagonal structure (space group $R\bar{3}m$) \cite{Mueller}. Figure \ref{fig:structure1} shows the crystal structure of the hexagonal Cs$_2$Cu$_3$MF$_{12}$ family and its projection onto the $c$ plane. As shown in Fig. \ref{fig:structure1}(b), CuF$_6$ octahedra are linked in the $c$ plane sharing corners. Magnetic Cu$^{2+}$ ions with $S\,{=}\,1/2$ located at the center of the octahedra form a uniform Kagom\'{e} layer in the $c$ plane. The nearest neighbor exchange interactions are equivalent, because all the Cu$^{2+}$ ions are crystallographically equivalent. CuF$_6$ octahedra are elongated along the principal axes, which makes an angle of approximately 11$^{\circ}$ with the $c$ axis. Because the elongated axes of the octahedra are approximately parallel to the $c$ axis, the hole orbitals $d(x^2\,{-}\,y^2)$ of Cu$^{2+}$ spread in the Kagom\'{e} layer. The nearest neighbor superexchange interaction $J$ through F$^-$ ion in the Kagom\'{e} layer is antiferromagnetic and strong \cite{Yamabe}, because the bond angle of the superexchange $\mathrm{Cu^{2+}\,{-}\,F^{-}\,{-}\,Cu^{2+}}$ is approximately $140^{\circ}$. The ferromagnetic superexchange occurs only when the bond angle is close to $90^{\circ}$. The interlayer exchange interaction $J'$ should be much smaller than $J$, because magnetic Cu$^{2+}$ layers are sufficiently separated by nonmagnetic Cs$^+$, M$^{4+}$ and F$^-$ layers. Thus, the present hexagonal Cs$_2$Cu$_3$MF$_{12}$ family can be expected to be quasi-two-dimensional $S\,{=}\,1/2$ HKAF. However, these systems undergo structural phase transitions at $T_{\rm t}\,{\simeq}\,220$ and 170 K, respectively, and also magnetic phase transitions at $T_{\rm N}\,{\simeq}\,24$ K \cite{Yamabe}. In this study, we synthesized Cs$_2$Cu$_3$SnF$_{12}$ that has the same crystal structure as Cs$_2$Cu$_3$ZrF$_{12}$ and Cs$_2$Cu$_3$HfF$_{12}$, and performed precise magnetic susceptibility measurements on Cs$_2$Cu$_3$MF$_{12}$ with M\,{=}\,Zr, Hf and Sn using high quality single crystals. In this paper, we will report the results. As shown below, the high-temperature susceptibilities of these systems can be almost perfectly described using the theoretical calculations on $S\,{=}\,1/2$ HKAF.
\begin{figure*}
\begin{center}
\includegraphics[scale =0.55]{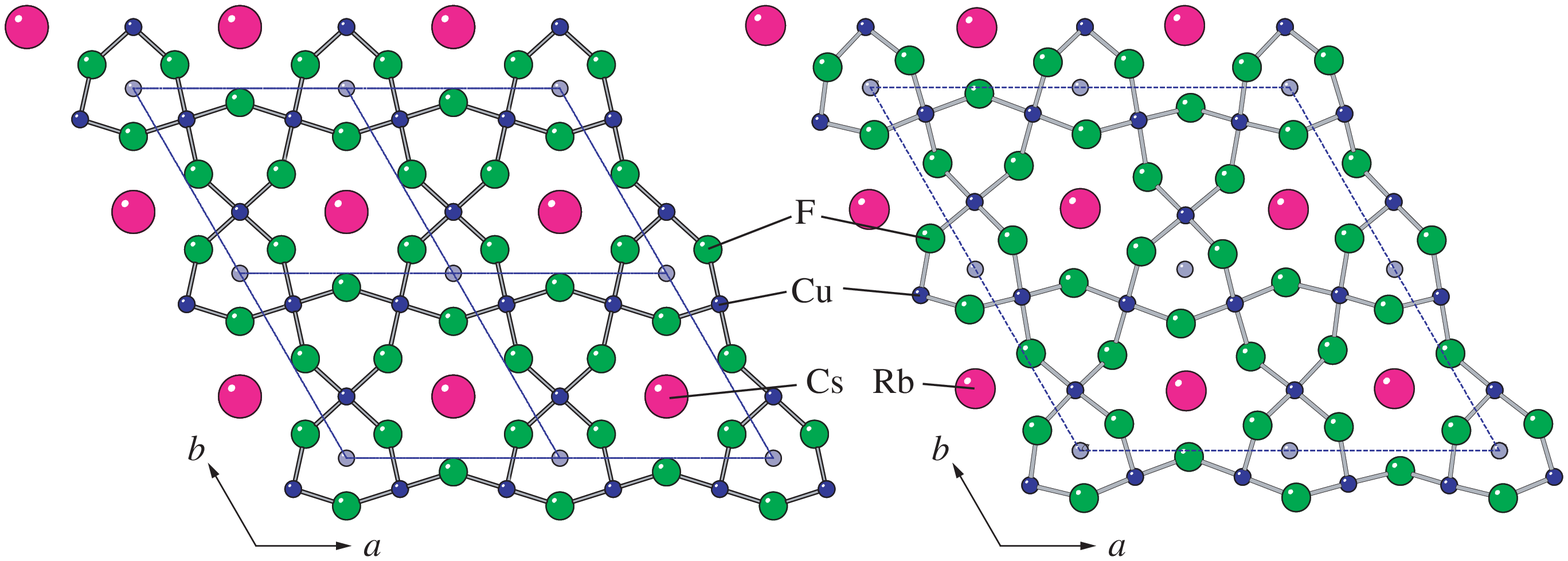}
\end{center}
\caption{Crystal structures of (a) Cs$_2$Cu$_{3}$SnF$_{12}$ and (b) Rb$_2$Cu$_{3}$SnF$_{12}$ viewed along the $c$ axis, where fluorine ions located outside the Kagom\'{e} layer are omitted. Dashed lines denote the unit cell.}
 \label{fig:structureCsRb}
 \end{figure*}
 
\begin{table}
\caption{Lattice constants $a$ and $c$ of Cs$_2$Cu$_3$MF$_{12}$ (M\,=\,Zr, Hf and Sn) and Rb$_2$Cu$_3$ZrF$_{12}$ in {\AA} unit.}
\label{table:1}
\begin{ruledtabular}
\begin{tabular}{ccccc} 
& M\,{=}\,Zr\,\footnote{Ref. \cite{Mueller}} & M\,{=}\,Hf\,\footnote{Ref. \cite{Mueller}} & M\,{=}\,Sn\,\footnote{Present work} & Rb$_2$Cu$_3$SnF$_{12}$\,\footnote{Ref. \cite{Morita}} \\ \hline
$a$ & 7.166 & 7.163 & 7.142(4) & 13.917(2) \\
$c$ & 20.46 & 20.49 & 20.381(14) & 20.356(3) \\
\end{tabular}
\end{ruledtabular}
\end{table}
\begin{table}
\caption{Fractional atomic coordinates and equivalent isotropic displacement parameters\,[$\rm{\AA}^2$] for Cs$_2$Cu$_3$SnF$_{12}$.}
\label{table:2}
\begin{ruledtabular}
\begin{tabular}{lcccc}
Atom  &  $x$  & $y$  & $z$ & $U_{\rm eq}$  \\ \hline
Cs & 0 & 0 & 0.1060(1) & 0.032(1)   \\
Cu & 0.5 & 0 & 0 & 0.011(1)  \\
Sn & 0 & 0 & 0.5 & 0.013(1)   \\
F(1) & 0.2042(2) & $-$0.2042(2) & 0.9845(1)  & 0.022(1)  \\
F(2) & 0.1312(2) & $-$0.1312(2) & 0.4465(1) & 0.036(1)  \\
\end{tabular}
\end{ruledtabular}
\end{table}

In the previous letter, we reported the crystal structure and magnetic properties of Rb$_2$Cu$_3$SnF$_{12}$ \cite{Morita}. Rb$_2$Cu$_3$SnF$_{12}$ has the hexagonal structure (space group $R{\bar 3}$), which is closely related to the structure of the above-mentioned Cs$_2$Cu$_{3}$MF$_{12}$. The unit cell in the Kagom\'{e} layer is enlarged to $2a\,{\times}\,2a$. Consequently, the Kagom\'{e} lattice is modified to have four types of neighboring exchange interaction. Figure \ref{fig:structureCsRb} shows a comparison of the crystal structures of Cs$_2$Cu$_{3}$SnF$_{12}$ and Rb$_2$Cu$_{3}$SnF$_{12}$. The structures are viewed along the $c$ axis, and fluorine ions located outside the Kagom\'{e} layer are omitted, so that magnetic Cu$^{2+}$ ions and exchange pathways are visible. From magnetic susceptibility and high-field magnetization measurements, it was found that the magnetic ground state of Rb$_2$Cu$_3$SnF$_{12}$ is a spin singlet with the triplet gap. In this paper, we will show the details of the analysis of the magnetic susceptibilities by exact diagonalization for a 12-site Kagom\'{e} cluster.
 
The arrangement of this paper is as follows: In section II, the experimental procedures and the crystal structure of newly synthesized Cs$_2$Cu$_3$SnF$_{12}$ are described. The results of the magnetic susceptibility measurements performed on Cs$_2$Cu$_{3}$MF$_{12}$ and Rb$_2$Cu$_3$SnF$_{12}$ are presented in section III. Analyses of the magnetic susceptibility data and discussion are also given in section III. Section IV is devoted to the conclusion. \\

\section{Experimental Details}
A$_2$Cu$_3$MF$_{12}$ crystals with M\,=\,Zr, Hf and Sn and A\,=\,Cs and Rb were synthesized in accordance with the chemical reaction $\mathrm{2AF + 3CuF_2 + MF_4}$ $\rightarrow$ $\mathrm{A_2Cu_3MF_{12}}$. 
$\mathrm{AF}$, $\mathrm{CuF_2}$ and $\mathrm{MF_4}$ were dehydrated by heating in vacuum at $\,{\sim}\,$100$^{\circ}$C. The materials were packed into a Pt tube of 9.6 mm inner diameter and $70{\sim}100$ mm length at the ratio of $3\,{:}\,3\,{:}\,2$. One end of the Pt tube was welded and the other end was tightly folded with pliers. Single crystals were grown from the melt. For Cs$_2$Cu$_3$ZrF$_{12}$ and Cs$_2$Cu$_3$HfF$_{12}$, the temperature of the furnace was lowered from 750 to 500$^{\circ}$C for four days, and from 800 to 550$^{\circ}$C for Cs$_2$Cu$_3$SnF$_{12}$ and Rb$_2$Cu$_3$SnF$_{12}$. Transparent colorless crystals are hexagonal-platelet shaped. 

The Cs$_2$Cu$_3$ZrF$_{12}$ and Cs$_2$Cu$_3$HfF$_{12}$ crystals obtained were identified by X-ray powder diffractions. The crystal data for Rb$_2$Cu$_3$SnF$_{12}$ have been reported in the previous paper \cite{Morita}. The crystal structure of Cs$_2$Cu$_3$SnF$_{12}$ has not been reported to date. Therefore, we analyzed the structural of Cs$_2$Cu$_3$SnF$_{12}$ at room temperature using a RIGAKU R-AXIS RAPID three-circle diffractometer equipped with an imaging plate area detector. Monochromatic Mo-K$\alpha$ radiation was used as an X-ray source. Data integration and global-cell refinements were performed using data in the range of $3.00^\circ\,{<}\,{\theta}\,{<}\,27.47^\circ$, and multi-scan empirical absorption correction was also performed. The total number of reflections observed was 2904. 290 reflections were found to be independent and 282 reflections were determined to satisfy the criterion $I\,{>}\,2{\sigma}(I)$. Structural parameters were refined by the full-matrix least-squares method using SHELXL-97 software. The final $R$ indices obtained were $R\,{=}\,0.0135$ and $wR\,{=}\,0.0325$. Cs$_2$Cu$_3$SnF$_{12}$ was found to be isostructural to Cs$_2$Cu$_3$ZrF$_{12}$ and Cs$_2$Cu$_3$HfF$_{12}$ (see Fig.\,\ref{fig:structure1}). The lattice constants of Cs$_2$Cu$_3$SnF$_{12}$ are listed in Table \ref{table:1} together with those of Cs$_2$Cu$_3$ZrF$_{12}$, Cs$_2$Cu$_3$HfF$_{12}$ and Rb$_2$Cu$_3$SnF$_{12}$. Fractional atomic coordinates and equivalent isotropic displacement parameters for Cs$_2$Cu$_3$SnF$_{12}$ are shown in Table \ref{table:2}. Lattice constants $a$ and $c$ are both smaller than those for Cs$_2$Cu$_3$ZrF$_{12}$ and Cs$_2$Cu$_3$HfF$_{12}$, which have almost the same lattice constants. 

The magnetic susceptibilities of the present four systems were measured in the temperature range of 1.8$-$400 K using a SQUID magnetometer (Quantum Design MPMS XL). High-field magnetization measurement on Rb$_2$Cu$_3$SnF$_{12}$ was performed by an induction method with a multilayer pulse magnet at the Institute for Solid State Physics, The University of Tokyo. Magnetic fields were applied parallel and perpendicular to the $c$ axis in both experiments.\\

\section{Results and Discussions}
\subsection{Cs$_2$Cu$_3$MF$_{12}$ (M\,=\,Zr, Hf and Sn)}
\begin{figure*}
\includegraphics[scale =0.48]{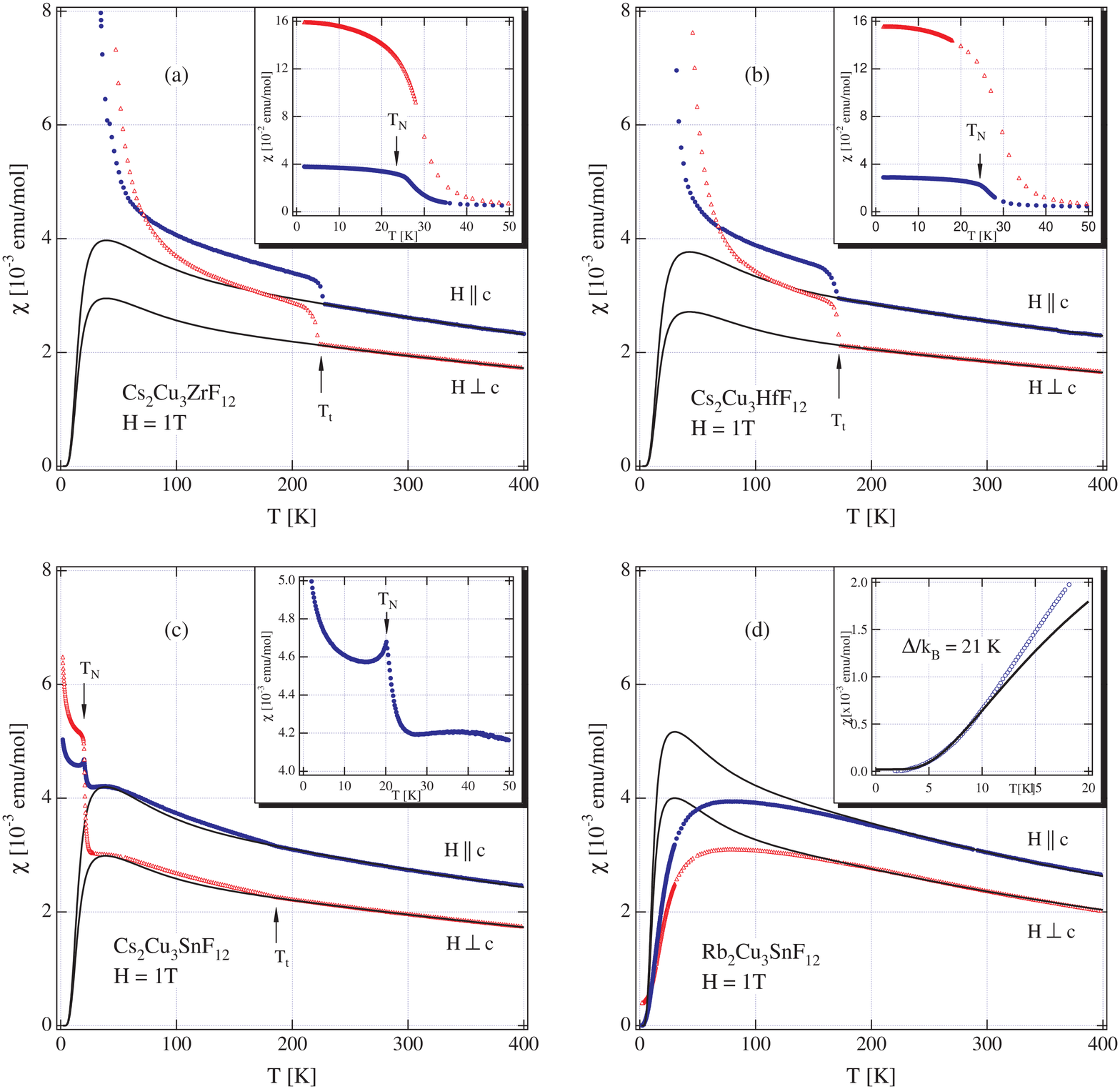}
\caption{Temperature dependences of magnetic susceptibilities of (a) Cs$_2$Cu$_3$ZrF$_{12}$, (b) Cs$_2$Cu$_3$HfF$_{12}$, (c) Cs$_2$Cu$_3$SnF$_{12}$ and (d) Rb$_2$Cu$_3$SnF$_{12}$ measured at $H\,{=}\,1$ T for $H\,{\parallel}\,c$ and $H\,{\perp}\,c$. Arrows in (a)$-$(c) denote structural and magnetic phase transition temperatures $T_{\rm t}$ and $T_{\rm N}$. Solid lines denote the fits using the theoretical susceptibilities for the uniform HKAF obtained from exact diagonalization for the 24-site Kagom\'{e} cluster \cite{Misguich2}. The magnetic parameters used for the fits for Cs compounds are given in Table \ref{table:Tt} and those for Rb$_2$Cu$_3$SnF$_{12}$ are shown in the text. Insets of (a)$-$(d) are low-temperature susceptibilities. The solid line in the inset of (d) denotes the fit using eq.\,(\ref{eq:chiRb}) with ${\Delta}/k_{\rm B}$=21\,K. }
\label{fig:sus1}
\end{figure*}

Cs$_2$Cu$_3$MF$_{12}$ (M\,=\,Zr, Hf and Sn) has the uniform Kagom\'{e} lattice at room temperature.
Figure \ref{fig:sus1} shows the temperature dependences of magnetic susceptibilities ${\chi}$ in these three systems. External magnetic field of $H\,{=}\,1$ T was applied for $H\,{\parallel}\,c$ and $H\,{\perp}\,c$. Note that one molar Cs$_2$Cu$_3$MF$_{12}$ contains three molar Cu$^{2+}$ ions. The susceptibility data were corrected for the diamagnetism ${\chi}_{\rm dia}$ of core electrons and the Van Vleck paramagnetism. The diamagnetic susceptibilities of individual ions were taken from the literature \cite{Selwood}. The Van Vleck paramagnetic susceptibility was calculated using
$
{\chi}_{\rm VV}^{\mu}\,{=}\,-(N{\mu}_{\rm B}^2/{\lambda}){\Delta g}_{\mu}\,{=}\,3.14\times 10^{-4}{\Delta g}_{\mu}\ {\rm emu/Cu^{2+}\,mol},
$
where ${\lambda}\,{=}\,-829$ cm$^{-1}$ is the coefficient of the spin-orbit coupling for Cu$^{2+}$ and ${\Delta g}_{\mu}\,{=}\,g_{\mu}-2$ is the anisotropy of the $g$ factor. To obtain the $g$ factors in the present systems, we performed electron spin resonance (ESR) measurements, but no spectrum was observed owing to the extremely large linewidth \cite{Linewidth}. Therefore, the $g$ factors and ${\chi}_{\rm VV}^{\mu}$ were determined self-consistently as ${\chi}_{\parallel c}/{\chi}_{\perp c}\,{=}\,(g_{\parallel c}/g_{\perp c})^2$ in the temperature range higher than the structural phase transition temperature $T_{\rm t}$. The $g$ factors obtained are listed in Table \ref{table:Tt}. These $g$ factors are consistent with $g_{\parallel}\,{=}\,2.48$ and $g_{\perp}\,{=}\,2.09$ measured in K$_2$CuF$_4$ and Rb$_2$CuF$_4$ with the K$_2$NiF$_4$ structure \cite{Sasaki}, where $g_{\parallel}$ and $g_{\perp}$ are the $g$ factors for the magnetic field parallel and perpendicular to the elongated axis of the CuF$_6$ octahedron, respectively.

With decreasing temperature from 400 K, the magnetic susceptibilities exhibit sudden jumps at $T_{\rm t}\,{=}\,225$ K and 172 K for Cs$_2$Cu$_3$ZrF$_{12}$ and Cs$_2$Cu$_3$HfF$_{12}$, respectively, while for Cs$_2$Cu$_3$SnF$_{12}$, the magnetic susceptibility displays a small bend anomaly at $T_{\rm t}\,{=}\,185$ K. The phase transition at $T_{\rm t}$ is attributed to structural phase transition. Because small hysteresis was observed at $T_{\rm t}$ for Cs$_2$Cu$_3$ZrF$_{12}$ and Cs$_2$Cu$_3$HfF$_{12}$, the structural phase transition is of first order. For Cs$_2$Cu$_3$SnF$_{12}$, the structural phase transition may be of second order, because the susceptibility anomaly at $T_{\rm t}$ is continuous. We performed the structural analyses below $T_{\rm t}$ on Cs$_2$Cu$_3$ZrF$_{12}$ and Cs$_2$Cu$_3$SnF$_{12}$ by X-ray diffraction, but we did not succeed in obtaining definite results. 
For Cs$_2$Cu$_3$SnF$_{12}$, it is certain that the unit cell in the Kagom\'{e} layer is enlarged to $2a\,{\times}\,2a$, as observed in Rb$_2$Cu$_3$SnF$_{12}$. However, it is unclear whether both structures are the same. For Cs$_2$Cu$_3$SnF$_{12}$, the value of 
${\eta}\,{=}\,({\chi}_{\parallel c}\,g_{\perp c}^2)/({\chi}_{\perp c}\,g_{\parallel c}^2)$ does not change at $T_{\rm t}$ and remains unity down to 40 K, where the $g_{\parallel c}$ and $g_{\perp c}$ are the $g$ factors for $T\,{>}\,T_{\rm t}$. This indicates that the angle between the elongated axis of CuF$_6$ octahedron and the $c$ axis is almost unchanged at $T_{\rm t}$. On the other hand, for Cs$_2$Cu$_3$ZrF$_{12}$ and Cs$_2$Cu$_3$HfF$_{12}$, the ${\eta}$ value decreases just below $T_{\rm t}$ to ${\eta}\,{=}\,0.88$ and 0.87, respectively. This implies that the angle between the elongated axis of CuF$_6$ octahedron and the $c$ axis increases below $T_{\rm t}$ on average, i.e., $11^{\circ}\,{\rightarrow}\,28^{\circ}$ and 27$^{\circ}$ for Cs$_2$Cu$_3$ZrF$_{12}$ and Cs$_2$Cu$_3$HfF$_{12}$, respectively.  

\begin{table}[tb]
\caption{Structural and magnetic phase transition temperatures $T_{\rm t}$\,[K] and $T_{\rm N}$\,[K], and $g$ factors, exchange constant $J/k_{\rm B}$\,[K] and effective Weiss constant ${\Theta}_{\rm eff}$\,[K] for $T\,{>}\,T_{\rm t}$ in Cs$_2$Cu$_3$MF$_{12}$ (M\,=\,Zr, Hf and Sn).}
\label{table:Tt}
\begin{ruledtabular}
\begin{tabular}{lcccccc}
 & $T_{\rm t}$ & $T_{\rm N}$ & $g_{\parallel c}$ & $g_{\perp c}$ & $J/k_{\rm B}$ & $-{\Theta}_{\rm eff}$\\
\hline
Cs$_2$Cu$_3$ZrF$_{12}$ & 225 & 23.5 & 2.43 & 2.10 & 244 & 520\\
Cs$_2$Cu$_3$HfF$_{12}$ & 172 & 24.5 & 2.47 & 2.10 & 266 & 620\\
Cs$_2$Cu$_3$SnF$_{12}$ & 185 & 20.2 & 2.48 & 2.10 & 240 & 540\\
\end{tabular}
\end{ruledtabular}
\end{table}

With further decreasing temperature, the magnetic susceptibilities of Cs$_2$Cu$_3$ZrF$_{12}$ and Cs$_2$Cu$_3$HfF$_{12}$ increase below 50 K, which is indicative of magnetic ordering accompanied by small ferromagnetic moments. The magnetic phase transition temperatures for both systems are approximately the same and are $T_{\rm N}\,{=}\,23.5$ K and 24.5 K, respectively. Because the anomaly of the magnetic susceptibility at $T_{\rm N}$ is not sharp in Cs$_2$Cu$_3$ZrF$_{12}$ and Cs$_2$Cu$_3$HfF$_{12}$, these N\'{e}el temperatures were determined by specific heat measurements \cite{Yamabe}. For Cs$_2$Cu$_3$SnF$_{12}$, the magnetic susceptibility has a rounded maximum at $T_{\rm max}\,{\simeq}\,38$ K as shown in the inset of Fig.\,\ref{fig:sus1}(c), and exhibits a sharp cusp and a sharp increase at $T_{\rm N}\,{=}\,20.2$ K for $H\,{\parallel}\,c$ and $H\,{\perp}\,c$, respectively, indicative of three-dimensional (3D) magnetic ordering.

We compare our experimental results for $T\,{>}\,T_{\rm t}$ with the theoretical susceptibility for the $S\,{=}\,1/2$ uniform HKAF obtained by Misguich and Sindzingre\,\cite{Misguich2} from exact diagonalization for the 24-site Kagom\'{e} cluster. The solid lines in Fig.\,\ref{fig:sus1}(a)$-$(c) denote the fits using the theoretical susceptibilities with the exchange parameters and the $g$ factors given in Table \ref{table:Tt}. Errors of the exchange parameters and the $g$ factors are ${\pm}2$ K and ${\pm}0.01$, respectively. The theoretical results provide an almost perfect description of the magnetic susceptibilities obtained for $T\,{>}\,T_{\rm t}$. The magnitude of the effective Weiss constant ${\Theta}_{\rm eff}$ obtained from the ${\chi}^{-1}$ vs $T$ plot for $T_{\rm t}\,{<}\,400$ K is $2.1\,{\sim}\,2.3$ times as large as $J/k_{\rm B}$, as predicted by theory \cite{Rigol1,Rigol2}. It is also observed that $k_{\rm B}|{\Theta}_{\rm eff}|/J$ increases a little with decreasing $T_{\rm t}$, the lower limit of temperature range. For Cs$_2$Cu$_3$SnF$_{12}$, the agreement between theoretical and experimental susceptibilities is fairly good even below $T_{\rm t}$. As shown in the inset of Fig.\,\ref{fig:sus1}(c), the magnetic susceptibility exhibits a rounded maximum at $T_{\rm max}\,{\simeq}\,38$ K, which coincides with the theoretical result $T_{\rm max}\,{\simeq}\,(1/6)J/k_{\rm B}\,{=}\,40$ K \cite{Elstner,Misguich2}. This small $T_{\rm max}$ as compared with $J/k_{\rm B}$ is characteristic of highly frustrated $S\,{=}\,1/2$ HKAF. Because the anomaly of the susceptibility at $T_{\rm t}$ is small in Cs$_2$Cu$_3$SnF$_{12}$, the change in the exchange interaction at $T_{\rm t}$ should be small. Below 30 K, the experimental susceptibility deviates significantly from the theoretical susceptibility owing to the occurrence of 3D ordering at $T_{\rm N}\,{=}\,20.2$ K. The ratio of $J/k_{\rm B}$ to $T_{\rm N}$ is approximately 10 in the present three systems. This large $J/(k_{\rm B}T_{\rm N})$ indicates the good two-dimensionality and the presence of the strong frustration.

Magnetization curves exhibit hysteresis below $T_{\rm N}$ in the present three systems, which indicates that weak ferromagnetic moments appear below $T_{\rm N}$. The weak ferromagnetic moment $M_{\rm wf}$ per Cu$^{2+}$ for $H\,{\parallel}\,c$ is much smaller than that for $H\,{\perp}\,c$. The values of $M_{\rm wf}$ in Cs$_2$Cu$_3$ZrF$_{12}$ and Cs$_2$Cu$_3$HfF$_{12}$ are almost the same, and $M_{\rm wf}\,{\simeq}\,0.07\,{\mu}_{\rm B}$ and ${\simeq}\,0.01\,{\mu}_{\rm B}$ for $H\,{\perp}\,c$ and $H\,{\parallel}\,c$, respectively. The rapid increase in the magnetic susceptibility below 50 K in these two systems arises from the weak ferromagnetic moment. In Cs$_2$Cu$_3$SnF$_{12}$, $M_{\rm wf}\,{\simeq}\,0.015\,{\mu}_{\rm B}$ for $H\,{\perp}\,c$, and $M_{\rm wf}\,{\simeq}\,0$ for $H\,{\parallel}\,c$. The weak ferromagnetic moment in Cs$_2$Cu$_3$SnF$_{12}$ is much smaller than those in Cs$_2$Cu$_3$ZrF$_{12}$ and Cs$_2$Cu$_3$HfF$_{12}$. Because the temperature variations of the magnetic susceptibilities of Cs$_2$Cu$_3$ZrF$_{12}$ and Cs$_2$Cu$_3$HfF$_{12}$ are similar to each other, the origin of the weak moment should be the same in these two systems, whereas the weak moment in Cs$_2$Cu$_3$SnF$_{12}$ is attributable to a different origin.

\begin{figure}[tbp]
	\begin{center}
		\includegraphics[scale =0.65]{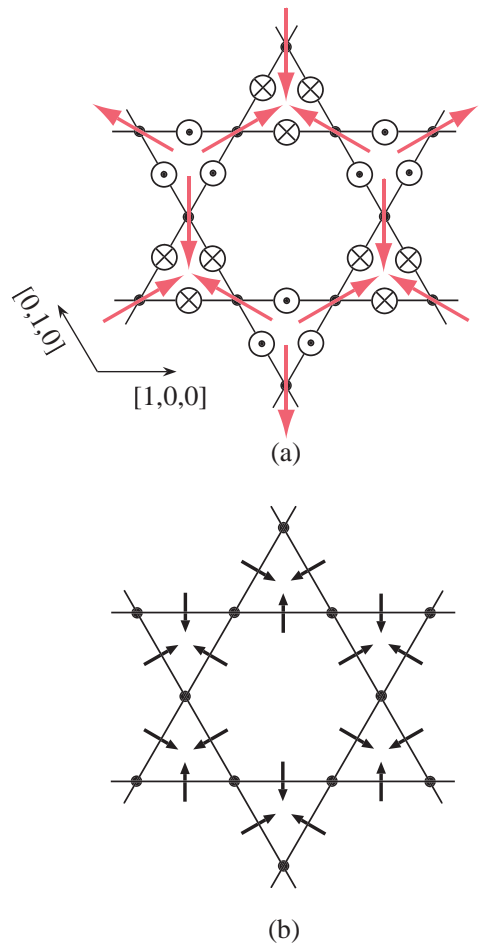}
	\end{center}
	\caption{Arrangement of the ${\bm D}$ vectors for $T\,{>}\,T_{\rm t}$ in $\mathrm{Cs_2Cu_3MF_{12}}$ systems, (a) the $c$ axis component $D^{\parallel}$ and (b) the $c$ plane component $D_{\perp}$. Symbols $\odot$ and $\otimes$ in (a) and arrows in (b) denote the local positive directions of parallel and perpendicular components $D^{\parallel}$ and $D^{\perp}$, respectively. Large arrows in (a) denote the $q\,{=}\,0$ spin structure stabilized when $D^{\parallel}\,{>}\,0$.}
	\label{fig:DM}
\end{figure}

\begin{figure}
	\begin{center}
		\includegraphics[scale =0.65]{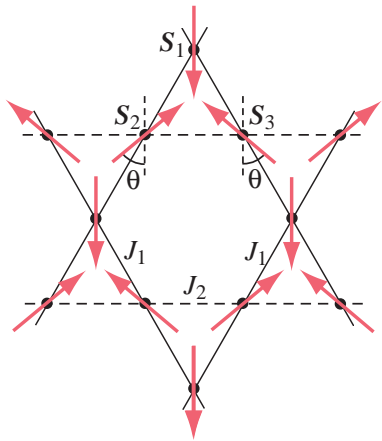}
	\end{center}
	\caption{Anisotropic exchange network composed of $J_1$ (solid lines) and $J_2$ (dashed lines), which results from orthorhombic distortion of the Kagom\'{e} lattice. Large arrows represent the $q\,{=}\,0$ spin structure composed of three sublattice spins $\bm S_1$, $\bm S_2$ and $\bm S_3$. $\theta$ denotes the angle between the $c$ plane component of $-\bm S_1$ and that of $\bm S_2$ or $\bm S_3$.}
	\label{fig:ortho}
\end{figure}

We consider two origins that give rise to the weak ferromagnetic moment. The first one is the DM interaction of the form ${\cal H}_{\rm DM}\,{=}\,\sum_{\langle i,j\rangle} {\bm D}_{ij}\,{\cdot}\,[{\bm S}_i\,{\times}\,{\bm S}_j]$. Because the details of the crystal structure below $T_{\rm t}$ are unclear, we first consider the configuration of the $\bm D$ vector in the high symmetric structure above $T_{\rm t}$. 
There is no inversion center at the middle point of two neighboring magnetic ions in the Kagom\'{e} lattice. This situation differs from that in the triangular lattice. Thus, in general, the DM interaction is allowed in the Kagom\'{e} lattice. There exists a mirror plane that passes the middle points of neighboring two Cu$^{2+}$ ions and is perpendicular to the line connecting these two ions. Therefore, the ${\bm D}$ vector should be parallel to the mirror plane \cite{Moriya}. Because there are two fold screw axes along the $[1, 0, 0], [0, 1, 0]$ and $[1, 1, 0]$ directions, the ${\bm D}$ vectors change their directions alternately along these directions. Thus, the configuration of the ${\bm D}$ vector should be as shown in Fig.\,\ref{fig:DM}. Two symbols $\odot$ and $\otimes$ in Fig.\,\ref{fig:DM}(a) and arrows in Fig.\,\ref{fig:DM}(b) denote the local positive directions of parallel and perpendicular components $D^{\parallel}$ and $D^{\perp}$, respectively. There is no a priori relation between $D^{\parallel}$ and $D^{\perp}$. The configuration of the ${\bm D}$ vector shown in Fig.\,\ref{fig:DM} is just the same as that discussed by Elhajal {\it et al.}\,\cite{Elhajal}.
The parallel component $D^{\parallel}$ acts as the easy-plane anisotropy that confines spins in the Kagom\'{e} layer and stabilizes the $q\,{=}\,0$ spin structure. The large arrows in Fig.\,\ref{fig:DM}(a) represent the $q\,{=}\,0$ spin structure that is stable for $D^{\parallel}\,{>}\,0$. Because the $c$ plane components of spins form the 120$^{\circ}$ structure, the parallel component $D^{\parallel}$ does not give rise to the weak moment parallel to the Kagom\'{e} layer.

The perpendicular component $D^{\perp}$ leads to the canting of spins from the Kagom\'{e} layer \cite{Elhajal}. The canting angle ${\varphi}$ is given by
\begin{eqnarray}
{\tan}\,2{\varphi}=\frac{2D^{\perp}}{\sqrt{3}J+D^{\parallel}}.
\label{eq:cant}
\end{eqnarray}
This spin canting produces the weak ferromagnetic moment of $M_{\rm wf}\,{=}\,g{\mu}_{\rm B}{\langle}S{\rangle}\,{\sin}\,{\varphi}$ perpendicular to the Kagom\'{e} layer. If the weak moments of neighboring layers do not cancel out, then weak net moment emerges along the $c$ axis. Although the details of the crystal structure below $T_{\rm t}$ is unclear at present, it is considered that the configuration of the $\bm D$ vector in the low-temperature crystal structure is close to that in the high-temperature crystal structure. Therefore, the weak ferromagnetic moment parallel to the $c$ axis observed in the ordered states of Cs$_2$Cu$_3$ZrF$_{12}$ and Cs$_2$Cu$_3$HfF$_{12}$ can be attributed to the perpendicular component of the $\bm D$ vector ($D^{\perp}$) for the DM interaction. In Cs$_2$Cu$_3$SnF$_{12}$, no weak moment was observed for $H\,{\parallel}\,c$. This should be because the weak moments of neighboring layers cancel out. Sharp cusp anomaly at $T_{\rm N}$ as shown in the inset of Fig.\,\ref{fig:sus1}(c) is often observed in jarosites \cite{Maegawa,Wills1,Inami1,Grohol3} and in the system where the weak moments cancel out \cite{Tanaka}.

As the second origin of the weak moment, we consider the orthorhombic distortion of the Kagom\'{e} lattice, as discussed by Wang {\it et al.} \cite{Wang}. This lattice distortion leads to two types of exchange interaction $J_1$ and $J_2$, as shown in Fig.\,\ref{fig:ortho}. If the 3D ordering occurs, the $120^{\circ}$ structure is modified to be isosceles. The angle $\theta$ between the $c$ plane component of $-\bm S_1$ and that of $\bm S_2$ or $\bm S_3$ is given by
\begin{eqnarray}
{\cos}\,{\theta}=\frac{J_1}{2J_2}.
\label{eq:ortho}
\end{eqnarray}
Thus, when the $q\,{=}\,0$ spin structure is realized, the weak moment of $M_{\rm wf}\,{=}\,g{\mu}_{\rm B}{\langle}S{\rangle}|2\,{\cos}\,{\theta}-1|/3$ is produced parallel to the Kagom\'{e} layer. Note that in the triangular lattice, the orthorhombic distortion of the lattice leads to the incommensurate helical spin structure \cite{Kato}, which is not accompanied by the weak moment. The weak moment of $M_{\rm wf}\,{\simeq}\,0.07\,{\mu}_{\rm B}$ observed in Cs$_2$Cu$_3$ZrF$_{12}$ and Cs$_2$Cu$_3$HfF$_{12}$ occurs when $J_1/J_2\,{\simeq}\,1.2$ or 0.8 with ${\langle}S{\rangle}\,{=}\,1/2$. Although the details of the low-temperature crystal structures of Cs$_2$Cu$_3$ZrF$_{12}$ and Cs$_2$Cu$_3$HfF$_{12}$ are unclear at present, it is probable that the hexagonal symmetry of the lattice is broken below $T_{\rm t}$ owing to the orthorhombic distortion, which leads to the anisotropic Kagom\'{e} lattice, as shown in Fig. \ref{fig:ortho}. We infer that the weak moment parallel to the Kagom\'{e} layer arises from such anisotropic exchange network. In Cs$_2$Cu$_3$SnF$_{12}$, the unit cell in the the Kagom\'{e} layer is enlarged to $2a\,{\times}\,2a$ below $T_{\rm t}$, but the lattice still has the hexagonal symmetry. Consequently, the exchange network is modified to be that for Rb$_2$Cu$_3$SnF$_{12}$ shown in Fig.\,\ref{fig:ExchangeRb}. In such exchange network that has the hexagonal symmetry, the weak moment does not appear. At present, we do not have clear explanation of the origin of the weak moment observed in Cs$_2$Cu$_3$SnF$_{12}$. It is possible that Cs$_2$Cu$_3$SnF$_{12}$ shows a domain structure below $T_{\rm t}$ and that the weak moment is induced in the domain boundaries where the hexagonal symmetry is locally broken. \\

\subsection{Rb$_2$Cu$_3$SnF$_{12}$}
\begin{figure}
	\begin{center}
		\includegraphics[scale =0.90]{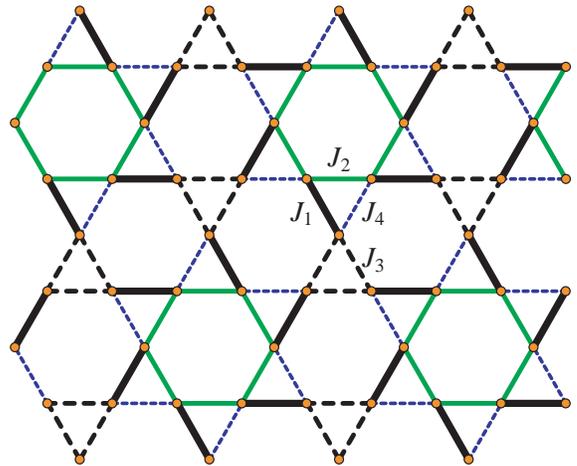}
	\end{center}
	\caption{Exchange network in the Kagom\'{e} layer of Rb$_2$Cu$_{3}$SnF$_{12}$. Pinwheels composed of $J_1$, $J_2$ and $J_4$ interactions are connected by triangles of $J_3$ interactions.}
	\label{fig:ExchangeRb}
\end{figure}

As mentioned above, the chemical unit cell of Rb$_2$Cu$_{3}$SnF$_{12}$ is enlarged to $2a\,{\times}\,2a$ in the Kagom\'{e} layer. In Fig.\,\ref{fig:structureCsRb}, we show a comparison of the crystal structures of Cs$_2$Cu$_{3}$SnF$_{12}$ and Rb$_2$Cu$_{3}$SnF$_{12}$. Figure \ref{fig:ExchangeRb} shows the exchange network in Rb$_2$Cu$_{3}$SnF$_{12}$. There are four types of nearest neighbor exchange interaction. Because the bond angle ${\alpha}$ of the exchange pathway Cu$^{2+}$\,$-$\,F$^{-}$\,$-$\,Cu$^{2+}$ ranges from ${\alpha}\,{=}\,123.9^{\circ}$ to 138.4$^{\circ}$, the exchange interactions should be of the same order as those of Cs$_2$Cu$_{3}$MF$_{12}$. The exchange interactions $J_1{\sim}J_4$ are labeled in decreasing order of ${\alpha}$, i.e., in decreasing order of magnitude.

Figure \ref{fig:sus1}(d) shows the temperature dependence of the magnetic susceptibilities of Rb$_2$Cu$_3$SnF$_{12}$ after the correction for the Curie-Weiss term owing to a small amount of impurities (${\simeq}0.3$\,\%), and for the diamagnetism of core electrons and the Van Vleck paramagnetism. Raw susceptibility data of Rb$_2$Cu$_3$SnF$_{12}$ are shown in the previous paper \cite{Morita}. With decreasing temperature, the magnetic susceptibilities of Rb$_2$Cu$_3$SnF$_{12}$ exhibit rounded maxima at $T_{\rm max}\,{\sim}\,70$ K and decrease rapidly. No magnetic ordering is observed. This result indicates clearly that the ground state is a singlet state with a triplet gap. The susceptibility for $H\,{\parallel}\,c$ is almost zero for $T\,{\rightarrow}\,0$, whereas that for $H\,{\perp}\,c$ is finite. 

In the present Kagom\'{e} family, elongated axes of CuF$_6$ octahedra incline alternately in the Kagom\'{e} layer, as shown in Fig.\,\ref{fig:structure1}(b). This leads to the staggered inclination of the principal axes of the $\bm g$ tensor. When an external field is applied, the staggered component of the $\bm g$ tensor produces the staggered field perpendicular to the external field \cite{Affleck}. The Zeeman interaction due to the staggered field and the DM interaction can have finite matrix elements between the singlet ground state and the excited triplet state, because they are antisymmetric with respect to the interchange of the interacting spins. Thus, we infer that the ground state has a small amount of triplet component through these antisymmetric interactions when subjected to the external field parallel to the Kagom\'{e} layer. This gives rise to the finite susceptibility at $T\,{=}\,0$ for $H\,{\perp}\,c$. 
\begin{figure}
	\begin{center}
		\includegraphics[scale =0.49]{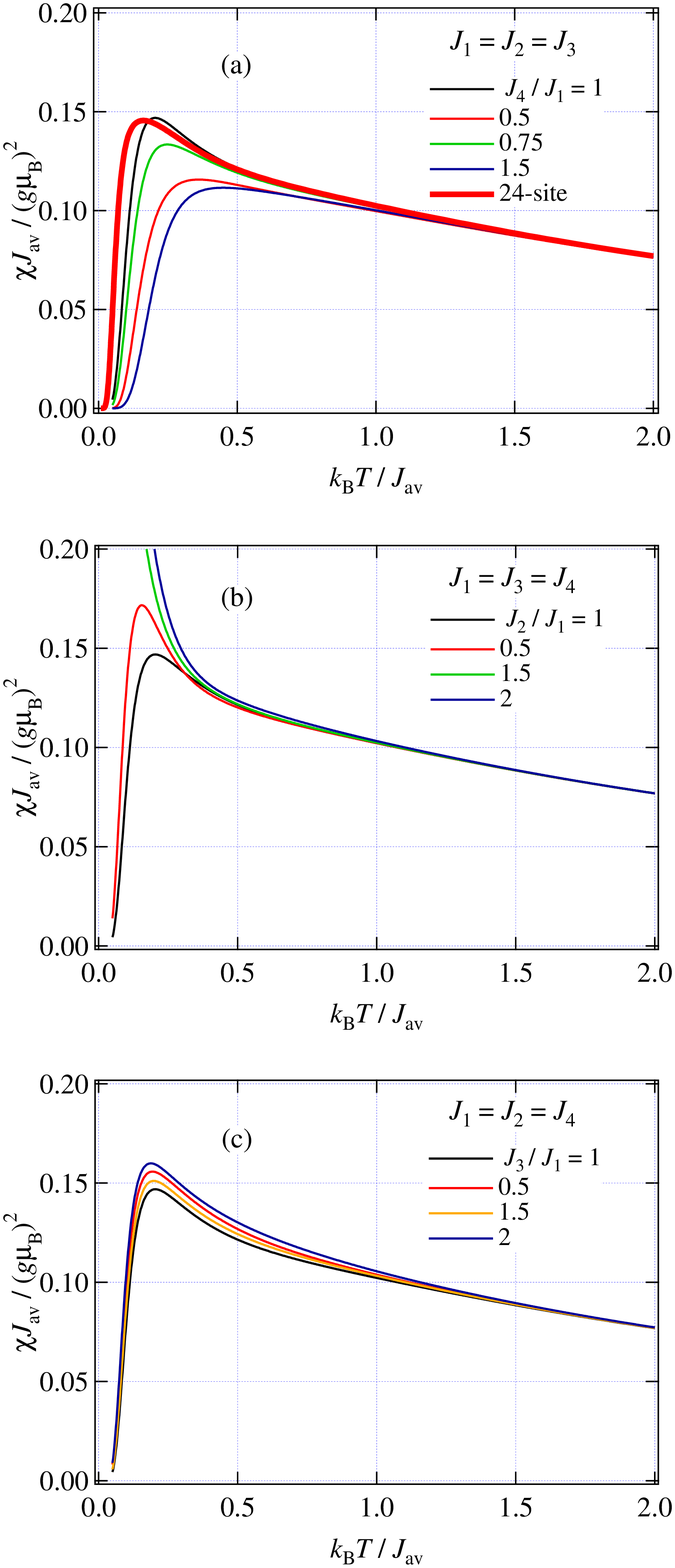}
	\end{center}
	\caption{Temperature dependence of magnetic susceptibilities per site obtained by exact diagonalizations for 12-site Kagom\'{e} cluster (a) for $J_4/J_1\,{=}\,0.5$, 0.75, 1 and 1.5 with $J_1\,{=}\,J_2\,{=}\,J_3$, (b) for $J_2/J_1\,{=}\,0.5$, 1, 1.5 and 2 with $J_1\,{=}\,J_3\,{=}\,J_4$ and (c) for $J_3/J_1\,{=}\,0.5$, 1, 1.5 and 2 with $J_1\,{=}\,J_2\,{=}\,J_4$. Temperature is scaled using the average of the exchange interactions $J_{\rm av}$. The thick solid line in (a) denotes the susceptibilities for the uniform HKAF obtained by exact diagonalization for the 24-site Kagom\'{e} cluster \cite{Misguich2}. 
	}
	\label{fig:SusTh}
\end{figure}

\begin{figure}
\begin{center}
\includegraphics[scale =0.50]{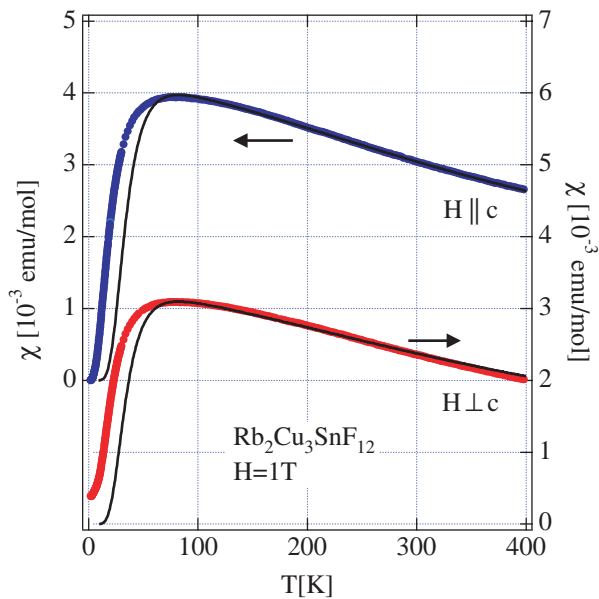}
\end{center}
\caption{Temperature dependence of the magnetic susceptibilities of Rb$_2$Cu$_3$SnF$_{12}$ measured for $H\,{\parallel}\,c$ and $H\,{\perp}\,c$. Solid lines denote the results obtained by exact diagonalization for a 12-site Kagom\'{e} cluster with the exchange parameters $J_1/k_{\rm B}\,{=}\,234$\,K, $J_2/k_{\rm B}\,{=}\,211$\,K, $J_3/k_{\rm B}\,{=}\,187$\,K and $J_4/k_{\rm B}\,{=}\,108$\,K, and $g_{\parallel}\,{=}\,2.44$ and $g_{\perp}\,{=}\,2.15$.}
\label{fig:SusRb}
\end{figure}

As shown in Fig.\,\ref{fig:sus1}(d), the magnetic susceptibility for $T$\,$>$\,150 K agrees with the theoretical susceptibilities (solid lines) for the $S$=1/2 uniform HKAF obtained by exact diagonalization for the 24-site Kagom\'{e} cluster with $J/k_{\rm B}\,{=}\,187$\,K, and $g_{\parallel}\,{=}\,2.43$ and $g_{\perp}\,{=}\,2.13$ \cite{Misguich2}. However, for $T$\,$<$\,150 K, the agreement between experimental and theoretical susceptibilities is poor. The theoretical susceptibility exhibits a sharp rounded maximum at $T_{\rm max}\,{\simeq}\,(1/6)J/k_{\rm B}\,{\simeq}\,30$ K, whereas the experimental susceptibility exhibits a broad maximum at $T_{\rm max}\,{\sim}\,70$ K. This is because exchange interactions in the Kagom\'{e} layer are not uniform. 

Figure\,\ref{fig:ExchangeRb} shows the exchange network in the Kagom\'{e} layer of Rb$_2$Cu$_3$SnF$_{12}$. The network consists of pinwheels composed of $J_1$, $J_2$ and $J_4$ interactions, which are connected by triangles of $J_3$ interactions. One chemical unit cell in the Kagom\'{e} layer contains 12 spins. Then, we performed the exact diagonalization for the 12-site Kagom\'{e} cluster under the periodic boundary condition. First, we calculated the uniform case, $J_1\,{=}\,J_2\,{=}\,J_3\,{=}\,J_4\,{=}\,1$, to confirm the validity of our calculation. The result obtained is the same as that obtained by Elstner and Young \cite{Elstner} from the exact diagonalization for the 12-site Kagom\'{e} cluster, and close to the result for the 24-site Kagom\'{e} cluster obtained by Misguich and Sindzingre \cite{Misguich2}. The comparison between susceptibilities obtained from the exact diagonalization for the 24- and 12-site Kagom\'{e} clusters is shown in Fig.\,\ref{fig:SusTh}(a). Both susceptibilities coincide for $k_{\rm B}T/J_{\rm av}\,{>}\,0.5$, and a small difference between them is observed for $k_{\rm B}T/J_{\rm av}\,{<}\,0.5$. 

Figure\,\ref{fig:SusTh}(a) shows the susceptibilities calculated for $J_4/J_1\,{=}\,0.5$, 0.75, 1 and 1.5, where we set $J_1\,{=}\,J_2\,{=}\,J_3$. In Fig.\,\ref{fig:SusTh}, temperature and susceptibilities are scaled using the average of exchange interactions $J_{\rm av}\,{=}\,(1/4)\sum_i J_i$. Whether for $J_4/J_1\,{>}\,1$ or $J_4/J_1\,{<}\,1$, the maximum susceptibility ${\chi}_{\rm max}$ decreases and $T_{\rm max}$ shifts toward the high-temperature side. The susceptibilities calculated for $J_4/J_1\,{=}\,0.5$ and 1.5 are close to the susceptibility observed in Rb$_2$Cu$_3$SnF$_{12}$. We also investigated the effects of $J_2$ and $J_3$, setting the others equal. Figure\,\ref{fig:SusTh}(b) shows the susceptibilities calculated for $J_2/J_1\,{=}\,0.5$, 1, 1.5 and 2. For $J_2/J_1\,{<}\,1$, ${\chi}_{\rm max}$ increases and $T_{\rm max}$ decreases with decreasing $J_2$, so that the rounded peak sharpens increasingly. The spin state for $J_2/J_1\,{\rightarrow}\,0$ is of interest.
When $J_2/J_1\,{>}\,1$, the susceptibility increases like obeying the Curie law for $T\,{\rightarrow}\,0$. In this case, six spins coupled through $J_2$ on a hexagon form a singlet state and three spins coupled through $J_3$ on a triangle form an $S\,{=}\,1/2$ state. This doublet state gives rise to the Curie law at low temperatures. Figure\,\ref{fig:SusTh}(c) shows the susceptibilities calculated for $J_3/J_1\,{=}\,0.5$, 1, 1.5 and 2 with $J_1\,{=}\,J_2\,{=}\,J_4$. The susceptibility is not largely influenced by $J_3$. 

When the hole orbitals $d(x^2\,{-}\,y^2)$ of neighboring Cu$^{2+}$ ions are linked through the $p$ orbital of F$^{-}$ ion as in the present systems, the antiferromagnetic exchange interaction becomes stronger with increasing bonding angle ${\alpha}$ of the exchange pathway Cu$^{2+}\,{-}\,$F$^{-}\,{-}\,$Cu$^{2+}$. It should be noted that the magnitude of the exchange constant for ${\alpha}\,{=}\,180^{\circ}$ was obtained to be $J/k_{\rm B}\,{\simeq}\,390$ K in KCuF$_3$\,\cite{Tennant} and K$_3$Cu$_2$F$_7$\,\cite{Manaka}. The bonding angle ${\alpha}_i$ for the exchange interaction $J_i$ in Rb$_2$Cu$_3$SnF$_{12}$ is as follows: ${\alpha}_1\,{=}\,138.4^{\circ}$, ${\alpha}_2\,{=}\,136.4^{\circ}$, ${\alpha}_3\,{=}\,133.4^{\circ}$, and ${\alpha}_4\,{=}\,123.9^{\circ}$. Hence, the condition $J_1\,{>}\,J_2\,{>}\,J_3\,{>}\,J_4$ must be realized. Varying exchange parameters under this condition, we calculated susceptibility. The best fit for $T\,{>}\,T_{\rm max}\,{\sim}\,70$ K is obtained using $J_1/k_{\rm B}\,{=}\,234(5)$\,K, $J_2/k_{\rm B}\,{=}\,211(5)$\,K, $J_3/k_{\rm B}\,{=}\,187(5)$\,K, and $J_4/k_{\rm B}\,{=}\,108(5)$\,K with $g_{\parallel}$=2.44(1) and $g_{\perp}$=2.15(1). The average of the exchange constants is $J_{\rm av}\,{=}\,185$ K. Solid lines in Fig.\,\ref{fig:SusRb} denote the susceptibilities calculated with these parameters. The magnitudes of these exchange interactions are valid from the fact that $J/k_{\rm B}\,{=}\,240$\,K and ${\alpha}\,{=}\,140.1^{\circ}$ in Cs$_2$Cu$_3$SnF$_{12}$, and $J/k_{\rm B}\,{=}\,103$\,K and ${\alpha}\,{=}\,129.1^{\circ}$ in KCuGaF$_6$ \cite{Morisaki}. In these two antiferromagnets, the exchange interactions were determined with high accuracy. For $T\,{<}\,T_{\rm max}$, the calculated susceptibility decreases more rapidly than the experimental susceptibility because of the finite-size effect. Calculation for a larger Kagom\'{e} cluster may give a better description of low-temperature susceptibility.

The exponential temperature variation of the low-temperature susceptibility indicates the presence of the triplet gap. The asymptotic low-temperature susceptibility for a 2D gapped system with a parabolic dispersion above the gap ${\Delta}$ is expressed as \cite{Troyer,Stone}
\begin{eqnarray}
{\chi}=A\,{\exp}\left(-{\Delta}/k_{\rm B}T\right).
\label{eq:chiRb}
\end{eqnarray}
The magnitude of the gap in Rb$_2$Cu$_3$SnF$_{12}$ is estimated to be ${\Delta}/k_{\rm B}\,{=}\,21(1)$\,K by fitting eq.\,(\ref{eq:chiRb}) to the low-temperature susceptibility data for $H{\parallel}c$. The solid line in the inset of Fig. \ref{fig:sus1}(d) denotes the fit.
\begin{figure}
\begin{center}
\includegraphics[scale =0.48]{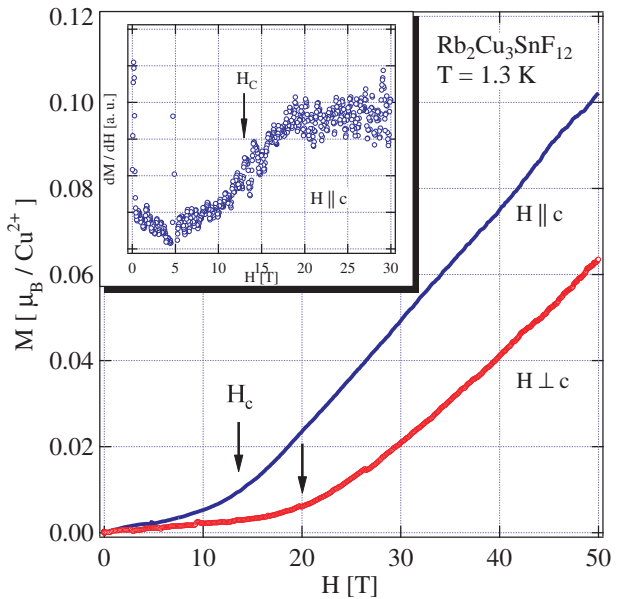}
\end{center}
\caption{Magnetization curves of Rb$_2$Cu$_3$SnF$_{12}$ measured at $T$=1.3 K for $H{\parallel}c$ and $H{\perp}c$. Arrows indicate the critical field $H_{\rm c}$. The inset shows $dM/dH$ vs $H$ for $H{\parallel}c$.}
\label{fig:MH}
\end{figure}

We also performed high-field magnetization measurements to evaluate the triplet gap directly. Figure \ref{fig:MH} shows magnetization curves and $dM/dH$ vs $H$ measured at $T$=1.3 K for $H{\parallel}c$ and $H{\perp}c$. The magnetization is small up to the critical field $H_{\rm c}$ indicated by arrows and increases rapidly. This magnetization behavior is typical of the gapped spin system. Level crossing between the ground and excited states occurs at $H_{\rm c}$. The bend anomaly at $H_{\rm c}$ is not sharp but is rather rounded. We infer that the smearing of the transition at $H_{\rm c}$ is ascribed to the staggered Zeeman and DM interactions that are antisymmetric with respect to the interchange of spins. We assign the critical field $H_{\rm c}$ to the field of inflection in $dM/dH$ as shown in the inset of Fig.\,\ref{fig:MH}. The critical fields obtained for $H{\parallel}c$ and $H{\perp}c$ are $H_{\rm c}\,{=}\,13(1)$\,T and 20(1)\,T, respectively. These critical fields do not coincide when normalized with the $g$-factor as $(g/2)H_{\rm c}$. Because the magnetic susceptibility for $H{\perp}c$ is finite even at $T\,{=}\,0$, the ground state has a finite triplet component. Thus, the ground state energy is no longer independent of the external field but decreases with the external field. This leads to an increase in the critical field. Therefore, we evaluate the gap to be ${\Delta}/k_{\rm B}\,{=}\,21(1)$\,K from $H_{\rm c}\,{=}\,13(1)$\,T obtained for $H{\parallel}c$. The magnitudes of the gap evaluated from the critical field and the low-temperature susceptibility coincide. It should be noted that ${\Delta}\,{=}\,0.11J_{\rm av}$ observed in Rb$_2$Cu$_3$SnF$_{12}$ and the triplet gap of ${\Delta}\,{\sim}\,0.1J$ predicted for the $S\,{=}\,1/2$ uniform HKAF\,\cite{Waldtmann,Singh2} are of the same order of magnitude.

It is evident from the present measurements that the ground state of Rb$_2$Cu$_3$SnF$_{12}$ is a spin singlet with a triplet gap, while Cs$_2$Cu$_3$SnF$_{12}$ has an ordered ground state. Then, we discuss the cause of the different ground states in these systems.  Although the crystal lattice of Cs$_2$Cu$_3$SnF$_{12}$ is enlarged to $2a\,{\times}\,2a$ below the structural phase transition at $T_{\rm t}\,{=}\,185$ K, the deviation of the magnetic susceptibility from the theoretical susceptibility for the uniform HKAF is small down to 30 K. This implies that the dispersion of the exchange constant ($J_1\,{\sim}\,J_4$) in Cs$_2$Cu$_3$SnF$_{12}$ is smaller than that in Rb$_2$Cu$_3$SnF$_{12}$. The triplet gap produced in a Kagom\'{e} layer decreases with decreasing dispersion of the exchange constant. Conversely, the gap increases with increasing degree of dimerization. Therefore, the gap produced in a Kagom\'{e} layer should be smaller in Cs$_2$Cu$_3$SnF$_{12}$ than in Rb$_2$Cu$_3$SnF$_{12}$, and the gap in Cs$_2$Cu$_3$SnF$_{12}$ collapses more easily when weak interlayer interaction operates. Because the distances between Kagom\'{e} layers in Cs$_2$Cu$_3$SnF$_{12}$ and Rb$_2$Cu$_3$SnF$_{12}$ are almost the same, the interlayer exchange interactions in both systems are of the same order of magnitude. We infer that the difference between the gaps produced by the exchange interactions in the Kagom\'{e} layers of both systems is an important factor of the different ground states.

Recently, C\'{e}pas {\it et al.}\,\cite{Cepas} have discussed the effect of the DM interaction on the ground state of $S\,{=}\,1/2$ HKAF. The configuration of the $\bm D$ vectors that they assumed is the same as that shown in Fig.\,\ref{fig:DM}. Using exact diagonalization and finite-size scaling, they argued that with increasing parallel component $D^{\parallel}$, a quantum phase transition from a gapped state to an ordered state occurs, and that the quantum critical point is given by $D^{\parallel}_{\rm c}\,{\simeq}\,0.1J$. In the present systems, the magnitude of the $\bm D$ vector is roughly estimated to be $D\,{\sim}\,({\Delta}g/g)J\,{\simeq}\,0.2J$ \cite{Moriya} with ${\Delta}g\,{\simeq}\,0.45$. Because ESR spectrum was not observed because of the extremely large linewidth produced by the large DM interaction \cite{Linewidth}, the ratio of $D^{\parallel}$ to $D^{\perp}$ is unclear. However, it is probable that $D^{\parallel}$ exceeds $0.1J$, so that the ground state is gapless for triplet excitations. In this case, 3D ordering occurs at finite temperature with the help of interlayer interactions, as observed in Cs$_2$Cu$_3$SnF$_{12}$. We infer that the difference in magnitude of the parallel component of the $\bm D$ vector is the second factor leading to the different ground state between Cs$_2$Cu$_3$SnF$_{12}$ and Rb$_2$Cu$_3$SnF$_{12}$.\\


\section{Conclusions}
We have presented the results of magnetic susceptibility measurements on the single crystals of Cs$_2$Cu$_3$MF$_{12}$ (M=Zr, Hf and Sn), which are described as $S\,{=}\,1/2$ HKAF. In these systems, structural phase transitions were observed at $T_{\rm t}\,{=}\,225$ K, 172 K  and 185 K for M\,=\,Zr, Hf and Sn, respectively. The magnetic susceptibilities observed for $T\,{>}\,T_{\rm t}$ are almost perfectly described using theoretical susceptibilities obtained by exact diagonalization for the 24-site Kagom\'{e} cluster with the exchange parameters and the $g$ factors shown in Table\,\ref{table:Tt}. For Cs$_2$Cu$_3$SnF$_{12}$, the agreement between theoretical and experimental susceptibilities is fairly good even below $T_{\rm t}$. The magnetic susceptibility exhibits a rounded maximum at $T_{\rm max}\,{\simeq}\,38$ K, which coincides with the theoretical result $T_{\rm max}\,{\simeq}\,(1/6)J/k_{\rm B}$.
Magnetic ordering accompanied by the weak ferromagnetic moment occurs at $T_{\rm N}\,{=}\,23.5$ K, 24.5 K and 20.0 K for M=Zr, Hf and Sn, respectively. The origins of the weak ferromagnetic moment should be ascribed to the lattice distortion that breaks the hexagonal symmetry of the exchange network for $T\,{<}\,T_{\rm t}$ and the DM interaction. 

We have also presented the results of magnetic measurements on Rb$_2$Cu$_3$SnF$_{12}$, which is described as a modified Kagom\'{e} antiferromagnet with four types of exchange interaction. The results of magnetic susceptibility and high-field magnetization measurements revealed that the ground state is a spin singlet with a triplet gap. Using exact diagonalization for a 12-site Kagom\'{e} cluster, we analyzed the magnetic susceptibility and evaluated individual exchange interactions. We have discussed the causes leading to the different ground states in Cs$_2$Cu$_3$SnF$_{12}$ and Rb$_2$Cu$_3$SnF$_{12}$. We infer that the difference in the dispersion of the exchange constant in the Kagom\'{e} layer and/or in the magnitude of the parallel component of the $\bm D$ vector causes the different ground states in these two systems. \\


\begin{acknowledgments}
We express our sincere thanks to G. Misguich for showing us his theoretical calculations of magnetic susceptibility. This work was supported by a Grant-in-Aid for Scientific Research (A) from the Japan Society for the Promotion of Science, and by a Global COE Program ``Nanoscience and Quantum Physics'' at Tokyo Tech and a Grant-in-Aid for Scientific Research on Priority Areas ``High Field Spin Science in 100 T'' both funded by the Japanese Ministry of Education, Culture, Sports, Science and Technology. 
\end{acknowledgments}



\begin{thebibliography}{99} 
\bibitem{Harrison} 
A. Harrison, J. Phys.: Condens. Matter \textbf{16}, S553 (2004).
\bibitem{ML} 
G. Misguich and C. Lhuillier, {\it Frustrated Spin Systems}, ed. H. T. Diep (World Science, Singapore, 2005) p. 229.
\bibitem{Harris} 
A. B. Harris, C. Kallin, and A. J. Berlinsky, Phys. Rev. B \textbf{45}, 2899 (1992).
\bibitem{Chubukov} 
A. Chubukov, Phys. Rev. Lett. \textbf{69}, 832 (1992).
\bibitem{Reimers} 
J. N. Reimers and A. J. Berlinsky, Phys. Rev. B \textbf{48}, 9539 (1993).
\bibitem{Zeng1} 
C. Zeng and V. Elser, Phys. Rev. B \textbf{42}, 8436 (1990).
\bibitem{Sachdev} 
S. Sachdev, Phys. Rev. B \textbf{45}, 12377 (1992).
\bibitem{Chalker} 
J. T. Chalker and J. F. G. Eastmond, Phys. Rev. B \textbf{46}, 14201 (1992).
\bibitem{Elstner} 
N. Elstner and A. P. Young, Phys. Rev. B \textbf{50}, 6871 (1994).
\bibitem{Nakamura} 
T. Nakamura and S. Miyashita, Phys. Rev. B \textbf{52}, 9174 (1995).
\bibitem{Zeng2} 
C. Zeng and V. Elser, Phys. Rev. B \textbf{51}, 8318 (1995).
\bibitem{Hida1} 
K. Hida, J. Phys. Soc. Jpn. \textbf{69}, 4003 (2000).
\bibitem{Hida2} 
K. Hida, J. Phys. Soc. Jpn. \textbf{70}, 3673 (2001).
\bibitem{Lecheminant} 
P. Lecheminant, B. Bernu, C. Lhuillier, L. Pierre, and P. Sindzingre, Phys. Rev. B \textbf{56}, 2521 (1997).
\bibitem{Waldtmann} 
Ch. Waldtmann, H.-U. Everts, B. Bernu, C. Lhuillier, P. Sindzingre, P. Lechminant, and L. Pierre, Eur. Phys. J. B \textbf{2}, 501 (1998).
\bibitem{Mila} 
F. Mila, Phys. Rev. Lett. \textbf{81}, 2356 (1998).
 
\bibitem{Syromyatnikov}
A. V. Syromyatnikov and S. V. Maleyev, Phys. Rev. B \textbf{66}, 132408 (2002).
\bibitem{Jiang} 
H. C. Jiang, Z. Y. Weng, and D. N. Sheng, Phys. Rev. Lett. \textbf{101}, 117203 (2008).
\bibitem{Nikolic} 
P. Nikolic and T. Senthil, Phys. Rev. B \textbf{68}, 214415 (2003).
\bibitem{Budnik} 
R. Budnik and A. Auerbach, Phys. Rev. Lett. \textbf{93}, 187205 (2004).
\bibitem{Singh1}
R. R. P. Singh and D. A. Huse, Rev. B \textbf{76}, 180407(R) (2007).
\bibitem{Singh2}
R. R. P. Singh and D. A. Huse, Rev. B \textbf{77}, 144415 (2008).
\bibitem{Yang}
B. -J. Yang, Y. B. Kim, J. Yu and K. Park, Rev. B \textbf{77}, 224424 (2008).

\bibitem{Mambrini} 
M. Mambrini and F. Mila, Eur. Phys. J. B \textbf{17}, 651 (2000).
\bibitem{Hastings}
M. B. Hastings, Phys. Rev. B \textbf{63}, 014413 (2000).
\bibitem{Ryu} 
S. Ryu, O. I. Motrunich, J. Alicea, and M. P. A. Fisher, Phys. Rev. B \textbf{75}, 184406 (2007).
\bibitem{Ran} 
Y. Ran, M. Hermele, P. A. Lee, and X. -G. Wen, Phys. Rev. Lett. \textbf{98}, 117205 (2007).
\bibitem{Hermele}
M. Hermele, Y. Ran, P. A. Lee, and X. -G. Wen, Rev. B \textbf{77}, 224413 (2008).

\bibitem{Sindzingre} 
P. Sindzingre, G. Misguich, C. Lhuillier, B. Bernu, L. Pierre, Ch. Waldtmann, and H.-U. Everts, Phys. Rev. Lett. \textbf{84}, 2953 (2000).
\bibitem{Misguich} 
G. Misguich and B. Bernu, Phys. Rev. B \textbf{71}, 014417 (2005).
\bibitem{Misguich2} 
G. Misguich and P. Sindzingre, Eur. Phys. J. B \textbf{59}, 305 (2007).
\bibitem{Rigol1} 
M. Rigol and R. R. P. Singh, Phys. Rev. Lett. \textbf{98}, 207204 (2007).
\bibitem{Rigol2} 
M. Rigol and R. R. P. Singh, Phys. Rev. B \textbf{76}, 184403 (2007).

\bibitem{Takano1} 
M. Takano, T. Shinjo, M. Kiyama, and T. Takada, J. Phys. Soc. Jpn. \textbf{25}, 902 (1968).
\bibitem{Takano2} 
M. Takano, T. Shinjo, and T. Takada, J. Phys. Soc. Jpn. \textbf{30}, 1049 (1971).
\bibitem{Townsend}
M. G. Townsend, G. Longworth, and E. Roudaut, Phys. Rev. B \textbf{33}, 4919 (1986).
\bibitem{Maegawa} 
S. Maegawa, M. Nishiyama, N. Tanaka, A. Oyamada, and M. Takano, J. Phys. Soc. Jpn. \textbf{65}, 2776 (1996).
\bibitem{Wills1}
A. S. Wills, A. Harrison, C. Ritter, and R. I. Smith, Phys. Rev. B \textbf{61}, 6156 (2000).
\bibitem{Inami1} 
T. Inami, M. Nishiyama, S. Maegawa, and Y. Oka, Phys. Rev. B \textbf{61}, 12181 (2000).
\bibitem{Wills2}
A. S. Wills, Phys. Rev. B \textbf{63}, 064430 (2001).
\bibitem{Inami2}
T. Inami, T. Morimoto, M. Nishiyama, S. Maegawa, Y. Oka, and H. Okumura, Phys. Rev. B \textbf{64}, 054421 (2001). 
\bibitem{Wills3}
A. S. Wills, G. S. Oakley, D. Visser, J. Frunzke, A. Harrison, and K. H. Andersen, Phys. Rev. B \textbf{64}, 094436 (2001).
\bibitem{Morimoto} 
T. Morimoto, M. Nishiyama, S. Maegawa, and Y. Oka,  J. Phys. Soc. Jpn. \textbf{72}, 2085 (2003).
\bibitem{Grohol1}
D. Grohol, D. G. Nocera, and D. Papoutsakis, Phys. Rev. B \textbf{67}, 064401 (2003).
\bibitem{Nishiyama}
M. Nishiyama, S. Maegawa, T. Inami, and Y. Oka, Phys. Rev. B \textbf{67}, 224435 (2003).
\bibitem{Grohol3} 
D. Grohol, K. Matan, J. -H. Cho, S. -H. Lee, J. W. Lynn, D. G. Nocera, and Y. S. Lee, Nat. Mater. \textbf{4}, 323 (2005).
\bibitem{Leblanc} 
M. Leblanc, R. De Pape, G. Ferey, and J. Pannetier, Solid State Commun. \textbf{58}, 171 (1986).

\bibitem{Elhajal} 
M. Elhajal, B. Canals, and C. Lacroix, Phys. Rev. B \textbf{66}, 014422 (2002).

\bibitem{Wills4}
A. S. Wills, A. Harrison, S. A. M. Mentink, T. E. Mason, and Z. Tun, Europhys. Lett. \textbf{42}, 325 (1998).
\bibitem{Fak}
B. F{\aa}k, F. C. Coomer, A. Harrison, D. Visser, and M. E. Zhitomirsky, Europhys. Lett. \textbf{81}, 17006 (2008).

\bibitem{Wada} 
N. Wada, T. Kobayashi, H. Yano, T. Okuno, A. Yamaguchi, and K. Awaga, J. Phys. Soc. Jpn. \textbf{66}, 961 (1997).
\bibitem{Kambe}
Takashi Kambe, Y. Nogami, K. Oshima, W. Fujita, and K. Awaga, J. Phys. Soc. Jpn. \textbf{73}, 796 (2004).

\bibitem{Hiroi} 
Z. Hiroi, M. Hanawa, N. Kobayashi, M. Nohara, H. Takagi, Y. Kato, and M. Takigawa, J. Phys. Soc. Jpn. \textbf{70}, 3377 (2001).
\bibitem{Honda}
Z. Honda, K. Katsumata, and K. Yamada, J. Phys.: Condens. Matter \textbf{14}, L625 (2002).
\bibitem{Rogado} 
N. Rogado, M. K. Haas, G. Lawes, D. A. Huse, A. P. Ramirez, and R. J. Cava, J. Phys.: Condens. Matter \textbf{15}, 907 (2003).
\bibitem{Narumi}
Y. Narumi, K. Katsumata, Z. Honda, J. -C. Domenge, P. Sindzingre, C. Lhuillier, Y. Shimaoka, T. C. Kobayashi, and K. Kindo, Eur. Phys. Lett. \textbf{65}, 705 (2004).

\bibitem{Shores} 
M. P. Shores, E. A. Nytko, B. M. Bartlett, and D. G. Nocera, J. Am. Chem. Soc. \textbf{127}, 13462 (2005). 
\bibitem{Mendels} 
P. Mendels, F. Bert, M. A. de Vries, A. Olariu, A. Harrison, F. Duc, J. C. Trombe, J. S. Lord, A. Amato, and C. Baines, Phys. Rev. Lett. \textbf{98}, 077204 (2007).
\bibitem{Helton}
J. S. Helton, K. Matan, M. P. Shores, E. A. Nytko, B. M. Bartlett, Y. Yoshida, Y. Takano, A. Suslov, Y. Qiu, J. -H. Chung, D. G. Nocera, and Y. S. Lee, Phys. Rev. Lett. \textbf{98}, 107204 (2007).
\bibitem{Bert} 
F. Bert, S. Nakamae, F. Ladieu, D. L'H\^{o}te, P. Bonville, F. Duc, J. -C. Trombe, and P. Mendels, Phys. Rev. B \textbf{76}, 132411 (2007).
\bibitem{Lee} 
S. -H. Lee, H. Kikuchi, Y. Qiu, B. Lake, Q. Huang, K. Habicht, and K. Kiefer, Nat. Mater. \textbf{6}, 853 (2007).
\bibitem{Imai}
T. Imai, E. A. Nytko, B. M. Bartlett, M. P. Shores, and D. G. Nocera, Phys. Rev. Lett. \textbf{100}, 077203 (2008).
\bibitem{Olariu}
A. Olariu, P. Mendels, F. Bert, F. Duc, J. C. Trombe, M. A. de Vries, and A. Harrison, Phys. Rev. Lett. \textbf{100}, 087202 (2008).
\bibitem{Vries}
M. A. de Vries, K. V. Kamenev, W. A. Kockelmann, J. Sanchez-Benitez, and A. Harrison, Phys. Rev. Lett. \textbf{100}, 157205 (2008).
\bibitem{Zorko} 
A. Zorko, S. Nellutla, J. van Tol, L. C. Brunel, F. Bert, F. Duc, J.-C. Trombe, M. A. de Vries, A. Harrison, and P. Mendels, Phys. Rev. Lett. \textbf{101}, 026405 (2008).

\bibitem{Mueller} 
M. M\"{u}ller and B. G. M\"{u}ller, Z. Anorg. Allg. Chem. \textbf{621}, 993 (1995).
\bibitem{Yamabe}
Y. Yamabe, T. Ono, T. Suto, and H. Tanaka, J. Phys.: Condens. Matter \textbf{19}, 145253 (2007).
\bibitem{Morita}
K. Morita, M. Yano, T. Ono, H. Tanaka, K. Fujii, H. Uekusa, Y. Narumi, and K. Kindo, J. Phys. Soc. Jpn. \textbf{77} (2008) 043707.
\bibitem{Selwood} 
P. W. Selwood, {\it Magnetochemistry} (Interscience, New York, 1956) 2nd ed., Chap. 2, p. 78.

\bibitem{Linewidth} 
The origin of the ESR linewidth ${\Delta}H$ is considered to be the DM interaction. The linewidth is given by ${\Delta}H\,{\sim}\,D^2/(g{\mu}_{\rm B}J)$. The magnitude of the $\bm D$ vector is roughly given by $D\,{\sim}\,({\Delta}g/g)J$ \cite{Moriya}. Using ${\Delta}g\,{\simeq}\,0.45$ and $J/k_{\rm B}\,{\simeq}\,250$ K, we estimate the linewidth to be ${\Delta}H\,{\sim}\,10$ T. This linewidth is too large to be observed by conventional ESR technique.
\bibitem{Moriya} 
T. Moriya, Phys. Rev. \textbf{120}, 91 (1960).

\bibitem{Sasaki} 
S. Sasaki, N. Narita, and I. Yamada, J. Phys. Soc. Jpn. \textbf{64}, 2701 (1995).
\bibitem{Tanaka} 
H. Tanaka, F. Tsuruoka, T. Ishii, H. Izumi, K. Iio, and K. Nagata, J. Phys. Soc. Jpn. \textbf{55}, 2369 (1986).
\bibitem{Wang} 
F. Wang, A. Vishwanath, and Y. B. Kim, Phys. Rev. B \textbf{76}, 094421 (2007).
\bibitem{Kato} 
T. Kato, K. Iio, T. Hoshino, T. Mitsui, and H. Tanaka, J. Phys. Soc. Jpn. \textbf{61}, 275 (1992).

\bibitem{Affleck} 
I. Affleck and M. Oshikawa, Phys. B. \textbf{60}, 1038 (1999).
\bibitem{Tennant} 
D. A. Tennant, T. G. Perring, R. A. Cowley, and S. E. Nagler, Phys. Rev. Lett. \textbf{70}, 4003 (1993).
\bibitem{Manaka} 
H. Manaka, Y. Miyashita, Y. Watanabe, and T. Masuda, J. Phys. Soc. Jpn. \textbf{76} (2007) 044710.
\bibitem{Morisaki} 
R. Morisaki, T. Ono, H. Tanaka, and H. Nojiri, J. Phys. Soc. Jpn. \textbf{76} (2007) 063706.
\bibitem{Troyer} 
M. Troyer, H. Tsunetsugu, and D. W\"{u}rtz, Phys Rev. B \textbf{50}, 13515 (1994).
\bibitem{Stone} 
M. B. Stone, I. Zaliznyak, D. H. Reich, and C. Broholm, Phys. Rev. B \textbf{64} (2001) 144405.
 
\bibitem{Cepas} 
O. C\'{e}pas, C. M. Fong, P. W. Leung, and C. Lhuillier, Phys. Rev. B \textbf{78}, 140405(R) (2008).

\end{thebibliography}
\end{document}